\documentclass[preprint,12pt,authoryear]{elsarticle}

\usepackage{amssymb}
\usepackage{amsmath}

\journal{Artificial Intelligence in Medicine}

\usepackage[utf8]{inputenc} 
\usepackage[T1]{fontenc}    
\usepackage{hyperref}       
\usepackage{url}            
\usepackage{float}       
\usepackage{tabularx}       
\usepackage{multirow}       

\usepackage{booktabs}       
\usepackage{enumitem}       

\usepackage{amsmath, amsfonts}       
\usepackage{nicefrac}       
\usepackage{bm}             

\usepackage{microtype}      
\usepackage{lipsum}

\usepackage{natbib}         

\usepackage{tikz}           
\usetikzlibrary{shapes,arrows,decorations.markings, decorations.pathreplacing,patterns,calc}

\usepackage{epstopdf}
\usepackage{graphicx}
\graphicspath{ {Figures/} } 

\setcitestyle{numbers}
\begin{document}

\begin{frontmatter}



\title{A Bayesian network model for predicting cardiovascular risk}


\author[inst1]{J.M. Ordovas}

\affiliation[inst1]{organization={Nutrition and Genomics, JM-USDA-HNRCA, Tufts University},
            city={Boston},
            state={MASS},
            country={USA}}

\author[inst2]{D. Rios Insua}
\author[inst3,inst4]{A. Santos-Lozano}
\author[inst5,inst4]{A. Lucia}
\author[inst2]{A. Torres}
\author[inst6]{A. Kosgodagan}
\author[inst2]{J.M. Camacho}

\affiliation[inst2]{organization={ICMAT-CSIC},
            city={Madrid},
            country={Spain}}
\affiliation[inst3]{organization={i+Health, Dpt. Health Sciences, European University Miguel de Cervantes},
            city={Valladolid},
            country={Spain}}
   \affiliation[inst4]{organization={Physical Activity and Health Research Group, Inst. Inv. Sanitaria Hospital 12 de Octubre},
            city={Madrid},
            country={Spain}}         
    \affiliation[inst5]{organization={Fac. Sports Sciences, Universidad Europea de Madrid},
            city={Madrid},
            country={Spain}}        
        \affiliation[inst6]{organization={Inst. de Mathématiques Apliqueés, Université Catholique de
        L'Ouest},
            city={Angers},
            country={France}}      
\begin{abstract}
We propose a Bayesian network model to make inferences and predictions about cardiovascular risk. Both the structure and the probability tables in the underlying model are built using a large dataset collected in Spain from annual work health assessments, with uncertainty characterized through posterior distributions. We illustrate its use for public health practice, policy and research purposes. A freely available 
version of the software is included in an Appendix.
\end{abstract}




\end{frontmatter}




\section{Introduction}
Cardiovascular diseases (CVD) are the leading cause of mortality in Europe and are responsible for over 3.9 million deaths per year (45\% of all deaths). Additionally, the yearly costs for CVD treatment surpass 210 billion euros in Europe \citep{Wilkins2017}.

Prediction scores are important tools for the prevention and management of CVD. In the clinical setting, accurate individual prediction of CVD risk is highly desirable to support the screening of high-risk patients, guide clinicians to determine and develop appropriate intervention strategies, and increase patient adherence.

CVD epidemiology can be traced back to the 1930s, with early studies describing an increased prevalence of CVD-related mortality with industrialization and affluence. The still ongoing \emph{Framingham Heart Study} was founded in 1948 to examine the epidemiology of CVD and has provided a wealth of information on CVD risk factors (CVRFs) \citep{Mahmood2014}.
Such factors are defined as measurable characteristics associated with increased disease frequency and are viewed as significant independent predictors of an increased risk of presenting such disease. The World Health Organization (WHO) has established a set of \emph{modifiable} and \emph{non-modifiable} CVRFs, which mainly address myocardial infarction (heart attack) and stroke \citep{Kaptoge2019}. Non-modifiable factors are those over which individuals have no control, including \emph{genetics}, \emph{age}, \emph{sex}, and \emph{ethnicity}. By contrast, modifiable CVRFs comprise behavioral factors that may evolve with individual activity, including \emph{physical activity (PA)}, \emph{diet}, \emph{alcohol}, and \emph{tobacco consumption}. The list of modifiable CVRFs is complemented by health indicators directly influenced by individuals’ behavior, including \emph{body mass index}, \emph{blood pressure}, and \emph{lipidemia}, among others. It is worth noting that the list of CVRFs established by the WHO is subject to frequent revision and validation \citep{Benjamin2019,Wilkins2017}.

Interest in Bayesian networks (BNs) \cite{Jensen1996} in the healthcare community has increased over the last two decades \cite{Kyrimi2020}, \cite{kyrimi2021bayesian}, 
\cite{mclachlan2020bayesian}.
Their ability to consistently and systematically integrate large amounts of data makes them well-suited to quantify the conditional probability tables that BNs require. In particular, for CVD diagnosis and prognosis, BNs represent a natural set-up for analyzing dependence across variables concerning related risk factors for CVD. 
For instance, \cite{Farooq2012} developed a BN for coronary angiography risk assessment. Some years later,
\cite{Tylman2016} proposed a BN to perform real-time CVD diagnosis focusing on myocardial infarction, infective endocarditis and hyperpotassemia. \cite{Roberts2006} built a simple BN model of the CV system and evaluated its ability to predict unobservable variables regarding the internal patient state. In relation to CVD epidemiology, one of the earliest studies constructing a BN linking various CVRFs to estimate CVD risk was performed by \cite{Twardy2006}. Later, \cite{Thornley2013} constructed an epidemiological CVD BN using data collected on individuals free of CVD, with the purpose of following-up patients over two years and assessing CVD events. Likewise, \cite{Fuster-Parra2016} proposed a discrete BN to identify and analyze the relationships among thirteen epidemiological features in the heart domain.

The purpose of the present study was to analyze relationships between several well-established CVRFs and additional factors that may contribute to CVD, focusing on the positive impact of PA (see \cite{Ley,santos2021association} for earlier insights into the relevance of PA for CVD). To achieve this objective, we used a discrete BN, with each node representing a CVRF or a medical condition. Both the BN structure and its probability tables are learnt from a large dataset originating from annual health assessments. Importantly, we characterize the uncertainty about such probabilities through (higher-order) distributions.
While the typical objective in the aforementioned literature is the analysis of CVD risk scores, here we emphasize the predictive assessment of various CVRFs and further integrate additional factors such as \emph{depression}, \emph{sleep duration} or \emph{socioeconomic status}, and analyze them through a BN-based framework. One of the main advantages of our model is that it allows assessing predictions for each factor, or sets of factors, using probabilistic inference. This can be used as a complementary decision-support tool to facilitate diagnosis and treatment by allowing practitioners to better foresee patients’ conditions. Moreover, it can also serve as a validating tool to check how likely the patient’s condition could have been forecast by the model based on past data.

The paper is structured as follows: Section~\ref{sec:datase} describes the dataset used to structure and calibrate the BN; in section~ \ref{sec:Model}, we formulate the BN model describing the structure learning process as well as the marginal distribution for each node; section~\ref{sec:policy} presents several uses 
of the model; and, finally, section~\ref{sec:discussion} summarizes and discusses our findings.   An appendix briefly describes a freely available software to perform predictions
based on our proposed BN.

\section{Data set}\label{sec:datase}
Data were extracted from the annual work health assessments of adult workers affiliated with a private insurance provider in Spain, from 2012 to 2016 and were anonymized and secured. Data were complemented with census information based on postal code, allowing us to infer the associated socioeconomic status (combining occupation, economic activity and professional situation averaged across the entire postal code) and 
education level (level 0 to 4, where 0 represents illiterate people and 4 represents people with a university degree or higher).
We removed outliers, duplicates, misrecorded and missing values, and retained only the last assessment available for each individual leading to a total of 205,087 health assessments over the 4-year period.

The variables relevant in this study are listed in Table~\ref{tab:Variables} and are
grouped as follows:

\begin{itemize}
    \item Non-modifiable CVRFs: sex, age, education, and socioeconomic status.
    \item Modifiable CVRFs: body mass index (BMI), PA,
sleep duration, smoking profile, anxiety, and depression.

    \item Medical conditions: hypertension, hypercholesterolemia, and diabetes.

\end{itemize}
Table \ref{tab:Variables} provides the values 
of the variables. 
\noindent
\begin{table}[htbp]
    \centering
    \footnotesize
    \begin{tabular}{cccc}
       Variable    & Definition  & Levels  
       \\ [1ex] \hline \\ [-1ex]
       $ v_{1} $   & Sex  & \{\emph{female, male}\}   \\ 
       
       \multirow{2}{*}{$ v_{2} $}   & \multirow{2}{*}{Age}  & \{[18-24], (24-34], (34-44], \\
          &   &  (44-54], (54-64], (64-74]\}   \\
       
       $ v_{3} $   & Education level & \{1,2,3\}   \\
       
       $ v_{4} $   & Socioeconomic status  & \{1,2,3\}  \\
       
       $ v_{5} $   & Body mass index  & \{\emph{underw., normal, overw., obese}\}   \\
       
      $ v_{6} $   & Physical activity  & \{ \emph{insufficiently active (1), regularly active (2)} \}     \\
      
      $ v_{7} $   & Sleep duration  & \{ Short; Normal; 
      Excessive\}  \\
       
       $ v_{8} $   & Smoker profile  & \{\emph{non-smoker, ex-smoker, smoker}\}    \\
       
       $ v_{9} $   & Anxiety  & \{\emph{yes, no}\}    \\
       
       $ v_{10} $   & Depression & \{\emph{yes, no}\}   \\
       
        $ v_{11} $   & Hypertension  & \{\emph{yes, no}\}  \\
       
        $ v_{12} $   & Hypercholesterolemia  & \{\emph{yes, no}\}  \\
       
       $ v_{13} $   & Diabetes & \{\emph{yes, no}\}    \\
      
       
    \end{tabular}
    \caption{Variables in model}
    \label{tab:Variables}
\end{table}

Age was categorized into six age groups following the grouping of the Spanish National Statistical Institute in their National Sport Habits survey, as in \cite{Ley}: (18-24), (24-34), (34-44), (44-54), (54-64) and (64-74). Education and socioeconomic levels were discretized to 1, 2, 3 (larger indices for higher education or socioeconomic levels).

As for modifiable CVRFs, we followed the WHO BMI guidelines for
 {\em underweight} (<18.5 kg/m$^2$),
{\em normal weight} (18.5-25 kg/m$^2$),
{\em overweight} (25-29.9 kg/m$^2$),  and {\em  obese} ( $\geq$ 30 kg/m$^2$)
classes. Participants’ self-reported leisure-time PA levels were assessed as in \cite{santos2021association}, allowing the categorization of {\em insufficiently active} (not meeting WHO minimum recommendations for aerobic PA in adults, i.e., <150min/wk and <75min/wk in moderate and vigorous PA, respectively), or {\em regularly active} (meeting WHO guidelines).
Sleep duration was categorized as {\em short} (less than 6 hours), {\em normal} (6-9 hours) and {\em excessive} (> 9 hours).
The smoker profile indicated whether the patient was currently an active smoker, had never smoked or was an active smoker in the past. Finally, we extracted from the data whether the patient had been clinically diagnosed with anxiety and depression.

With regard to medical conditions, in addition to previous medical history,
we also determined whether patients had diabetes (i.e., medicated or glycaemia $\geq 125$ mg/dL), hypercholesterolemia (medicated or total cholesterol $\geq 240$ mg/dL), or hypertension (medicated or systolic/diastolic blood pressure $\geq 140$/$90$mm Hg).

 Table~\ref{tab:Post_Distr} lists the proportion of cases in various categories reflecting, by and large, the standard structure of the Spanish labor force (with the exception of the lower presence of females and of more senior people in our study than in the overall labor market due to the types of companies served by the insurance provider) and its health status. In particular, we found a high proportion of (i) overweight to obese subjects; (ii) smokers; (iii) hypertensive and hypercholesterolemic individuals; and (iv) a high proportion of subjects not meeting the WHO-recommended PA levels. By contrast, anxiety, diabetes and depression were at relatively low levels.

 {\small
\begin{table}[!htb]
   \centering
   \footnotesize
   \begin{tabularx}{\textwidth}{lllp{0.01cm}lll}
      \textbf{Variables} & \textbf{States} & \textbf{Marginal} & & \textbf{Variables} & \textbf{States} & \textbf{Marginal}
      \\ [1ex] \hline \\ [-1ex]
      BMI & Underweight & 1.16\% & & Diabetes & Yes & 2.51\% \\
       & Normal & 40.74\% & & & No & 97.49\% \\
       & Overweight & 40.31\% & & Hypertension & Yes & 15.05\% \\
       & Obese & 17.79\% & & & No & 84.95\% \\  
      Sex  & Female & 32.10\% & & Hypercholest. & Yes & 30.53\% \\
        & Male & 67.90\% & & & No & 69.47\% \\
      Smoker  & Non-Smoker & 50.14\% & & Physical Act. & 1 & 75.63\% \\
      profile & Ex-Smoker & 20.49\% & & & 2 & 24.37\% \\
       & Smoker & 29.37\% & & &   &   \\
      Age(y) & (18-24) & 0.71\% & & Educ. lev. & 1 & 0.21\% \\
       & (24-34) & 17.10 \% & & & 2 & 76.65\% \\
       & (34-44) & 36.81\% & & & 3 & 23.14\% \\
       & (44-54] & 30.37\% & & Sleep Dur. & < 6h & 11.66\% \\ 
       & (54-64) & 14.83\% & & & (6h-9h) & 88.25\% \\
       & (64-74) & 0.18\% & & & > 9h & 0.09\% \\
      Socioeconomic  & 1 & 37.83\% & & Depression & No & 99.52\% \\
       status & 2 & 35.15\% & & & Yes & 0.48\% \\
       & 3 & 27.02\% & & & & \\
      Anxiety & Yes & 2.68\% & & & & \\
       & No & 97.32\% & & & &
        \\ [1ex] \hline \\ [-1ex]
   \end{tabularx}
   \caption{Percentage of observations at each class}
   \label{tab:Post_Distr}
\end{table}
}

We performed this exploratory analysis for each of the years as a sensitivity analysis, revealing relatively minor differences. This suggested stability of the workforce health structure over the five years available, allowing us to safely aggregate the data over that period.
%

\section{A Bayesian network for cardiovascular risk factors}\label{sec:Model}

We built a discrete BN to estimate the joint distribution underlying the available data as the basis 
to make inferences and predictions on cases of relevance.
The variables in the BN nodes were coded as in Table~\ref{tab:Variables}. We used a two-stage search-and-score procedure to learn the structure of the BN.

First, we employed 
\emph{greedy thick thinning} \cite{Cheng97}, a Bayesian search algorithm to estimate the initial structure from the data, which is based on a conditional independence test that scores BN structures according to a conditional dependence measure. It is suitable when the amount of available data is large, as in our case, as it converges asymptotically to the eventual actual structure \citep{Pearl88}.

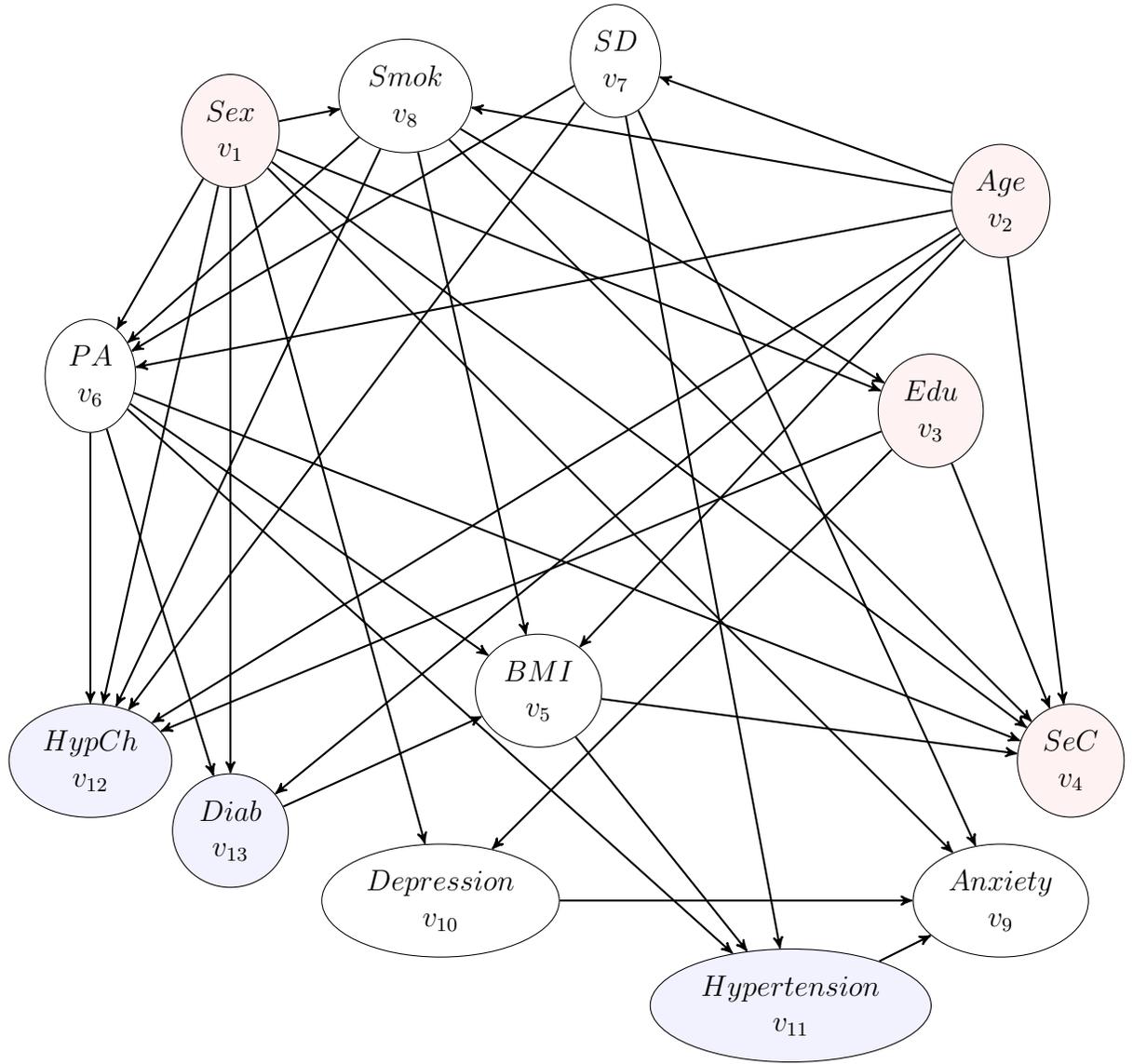
\begin{figure}[htbp]
	\centering
	\begin{tikzpicture}
	\tikzstyle{chance}=[ellipse, fill=red!5, draw,inner sep = 4pt, align = center]
	\tikzstyle{chance1}=[ellipse, fill=blue!5, draw,inner sep = 4pt, align = center]
	\tikzstyle{chance2}=[ellipse, draw,inner sep = 4pt, align = center]
	\tikzstyle{suite}=[->,>=stealth',thick]
	
	
	\node[chance2] (V1) at (1.4,-1) { $BMI $ \\ $ v_5 $};
	\node[chance] (V2) at (-3,7) {$Sex$ \\ $ v_1 $};
	\node[chance2] (V3) at (-.5,7.5) { $Smok$ \\ $ v_8 $ };
	\node[chance] (V4) at (8,6) {$Age$ \\ $ v_2 $};
	\node[chance] (V5) at (9,-2) {$SeC$ \\ $ v_4 $};
	\node[chance2] (V6) at (8,-4) {$Anxiety$ \\ $ v_9 $};
	\node[chance1] (V7) at (-3,-3) {$Diab$ \\ $ v_{13} $};
	\node[chance1] (V8) at (5,-5.5) {$ Hypertension$ \\ $ v_{11} $};
	\node[chance1] (V9) at (-5,-2) { $HypCh$ \\ $ v_{12} $};
	\node[chance2] (V10) at (-5,3.5) { $ PA $ \\ $ v_{6} $};
	\node[chance] (V11) at (7,3) { $ Edu $ \\ $ v_{3} $};
	\node[chance2] (V12) at (2.5,8) { $ SD $ \\ $ v_{7} $};
	\node[chance2] (V13) at (0,-4) { $ Depression $ \\ $ v_{10} $};
	
	
	\draw[suite] (V1) -- (V8); 
	\draw[suite] (V1) -- (V5); 
	
	\draw[suite] (V2) -- (V10); 
	\draw[suite] (V2) -- (V9); 
	\draw[suite] (V2) -- (V7); 
	\draw[suite] (V2) -- (V13); 
	\draw[suite] (V2) -- (V6); 
	\draw[suite] (V2) -- (V5); 
	\draw[suite] (V2) -- (V11); 
	\draw[suite] (V2) -- (V3); 
	
	\draw[suite] (V3) -- (V10); 
	\draw[suite] (V3) -- (V9); 
	\draw[suite] (V3) -- (V5); 
	\draw[suite] (V3) -- (V1); 
	\draw[suite] (V3) -- (V11); 
	
	\draw[suite] (V4) -- (V12); 
	\draw[suite] (V4) -- (V3); 
	\draw[suite] (V4) -- (V10); 
	\draw[suite] (V4) -- (V9); 
	\draw[suite] (V4) -- (V1); 
	\draw[suite] (V4) -- (V7); 
	\draw[suite] (V4) -- (V5); 
	
	
	
	\draw[suite] (V7) -- (V1); 
	
	\draw[suite] (V8) -- (V6); 
	
	
	\draw[suite] (V10) -- (V9); 
	\draw[suite] (V10) -- (V7); 
	\draw[suite] (V10) -- (V8); 
	\draw[suite] (V10) -- (V1); 
	\draw[suite] (V10) -- (V5); 
	
	\draw[suite] (V11) -- (V9); 
	\draw[suite] (V11) -- (V13); 
	\draw[suite] (V11) -- (V5); 
	
	\draw[suite] (V12) -- (V10); 
	\draw[suite] (V12) -- (V9); 
	\draw[suite] (V12) -- (V8); 
	\draw[suite] (V12) -- (V6); 
	
	\draw[suite] (V13) -- (V6); 

	\end{tikzpicture}		
	\caption{Layout of the Bayesian network structure. White (modifiable cardiovascular risk factors); light
	pink (non-modifiable cardiovascular risk factors); light blue (medical conditions). Abbreviations: $Edu$, education level; $SeC$, socioeconomic status ; $BMI$, body mass index; $PA$, physical activity; $SD$, sleep duration; $Smok$, smoker profile; $HypCh$, hypercholesterolemia; $Diab$: diabetes. }
	\label{fig:BN_Struct}	
\end{figure}

The second stage was an empirical iterative procedure involving several (three) experts in the CVD domain. The experts were shown various BN structures and asked to prune or add relevant edges reasoning on a cause-effect mechanism between “parent-children” sets of nodes. Various iterations were needed to agree on the final BN structure shown in Figure~\ref{fig:BN_Struct},
 with different colors for medical conditions (blue) and  modifiable (white) and non-modifiable (pink) CVRFs.\footnote{Note \cite{Jensen1996} that arcs connecting the structure are not to be interpreted in cause-effect terms between a parent and children. Indeed, as inference can be performed ’backwards’, inserted pieces of information may circulate both ways, as required for inference and prediction purposes. Thus, information-wise, parent and children nodes may be equally important whenever data is available for either of them.} 
As a consequence of the graphical
representation, the underlying suggested probabilistic model over the thirteen variables is

{\small 
\begin{align*}
    p (v_1, \dots ,v_{13}) & =
    \Big[ p(v_1) p(v_2) p(v_{3}| v_1, v_8) p(v_4|v_1,v_2,v_3,v_5,v_6,v_8) \Big] \times \\
  & \Big[ p(v_5|v_2, v_6, v_8 ) p(v_6|v_1, v_2, v_7, v_8 )
    p (v_7 | v_2)   p(v_8|v_1,v_2)   \\
    & p(v_9| v_1, v_7, v_{10}, v_{11}) p(v_{10}|v_1, v_3) \Big] \times \\
    & 
    \Big[ p(v_{11}| v_5, v_6, v_7 )  
    p(v_{12}|v_1, v_2, v_3, v_6, v_7 , v_8 ) 
    p(v_{13}|v_1, v_2, v_6) \Big] ,
   \end{align*}
}
\noindent where we have marked the three blocks of variables considered.


After defining the structure, we assessed the probability tables at various nodes using standard multinomial-Dirichlet models with uniform priors \citep{French}.  As an example, we provide the expected values of the distributions at node $v_7$, {\em sleep duration}, in Table \ref{kk}. Of note, the last column in the table suggests that 19.64\% of the population in this age group sleeps less than 6 hours and 80.33\% sleeps between 6 and 9 hours.

\begin{table}[htbp]
    \centering
    \footnotesize
    \begin{tabular}{ccccccc}\hline
       Sleep Duration (h)  /age(y)  & 18-24 & 24-34 & 34-44& 44-54 & 54-64 & 64-74 \\ \hline   
  $<6$   & 4.67 & 6.94 & 10.16 & 13.50 & 17.30 & 19.64 \\     
    $6-9$   & 94.98 & 92.93 & 89.77 & 86.43 & 82.61 & 80.33 \\   
     $>9$   & 0.35 & 0.13 & 0.07 & 0.07 & 0.09 & 0.03 \\ \hline
    \end{tabular}
    \caption{Probability table at node $v_7$. Probabilities as expected values of the corresponding posterior distributions are expressed as percentages.}
    \label{kk}
\end{table}
\noindent The proportion of workers with low sleep hours seems to increase with age. As an example, we tested whether the proportion of low sleeping hours for the 64–74 age group  (whose posterior
distribution is $\beta e (79,353)$) is larger than that for the 18–24 age group
(whose posterior distribution is $\beta e (59,1900)$),
with Bayesian hypothesis testing methods \citep{giron}.
 Our simulation showed that the probability for the first proportion being larger than the second is almost 100\%, supporting the above statement. On the other hand, when comparing the proportions for the 64–74 and 54–64 age groups, the posterior 
 probability was 75.7\%. Henceforth, we will use the point estimates of the probabilities based on posterior expectations. The model has been cross-validated, displaying proper generalization with unseen data \cite{nagarajan2013bayesian}. 
\section{Treatment practices and policy uses}\label{sec:policy}
We now describe several treatment practices and policy uses of the graphical structure, exploiting the ability of BNs to makes inferences, issue forecasts and support decisions: information, measurements and observations can be inserted and propagated throughout the network and modify the distributions in other nodes based on probabilistic computations using Bayes’ formula \cite{Jensen1996}, possibly applied several times. Accordingly, the network becomes a probabilistic database that summarizes the uncertainty about the CVD domain and may be queried to respond to diagnosis, treatment and policy issues.

\subsection{Diagnosis and evidence propagation}
First, we provide examples that combine inserted pieces of information to evaluate the predictive probability that a patient develops hypertension, a key point when dealing with CVD.

We first describe how such probabilities can be computed. These could refer to an individual with specific features for which we would forecast their likelihood of developing a particular medical condition, or to a group of individuals with such features for which we aim to predict the proportion of individuals that could develop such condition. For example, consider the scenario of a male aged over 44, sleeping less than 6 hours per night, overweight,  sedentary, and having anxiety. The conditional probability that this person has hypertension, given the previous information, is described through
{\small 
\begin{align*}
& Pr (v_{11} = y \, | \, v_1 = male, v_2 \geq 45, v_5 = overw., v_6 = 1 , 
v_7  = \{< 6h\}), v_9 = y ) = \\ \nonumber
& \frac { Pr (v_1 = male, v_2 \geq 45, v_5 = overw., v_6 = 1 , 
    v_7  = \{< 6h\}), v_9 = y, v_{11}= y )}
    { Pr (v_1 = male, v_2 \geq 45, v_5 = overw., v_6 = 1 , 
    v_7  = \{< 6h\}), v_9 = y )}   \stepcounter{equation}\tag{\theequation}\label{peterbrand} 
\end{align*}
}
 \noindent Simple but computationally intensive,
 calculations\footnote{All
 calculations performed with R.} based on the law of total probability would compute the probability of interest, which is 25.26\%, well above the marginal probability (15.05\%) of such an event shown in Table  \ref{tab:Post_Distr}.
This suggests an important increase in the probability of having hypertension for individuals with such a profile. In this case, had the individual not been diagnosed with such a medical condition, they should be notified about the elevated risk and the associated problems. Suppose now that the above calculations refer to a group with such features. Then, we would have identified a group in which to concentrate communication and intervention efforts, given the non-negligible increased probability of having the condition.

Table~\ref{tab:Prob_Queries}
lists other relevant examples, with a focus on the impact of PA. Specifically, using case \eqref{peterbrand} as a benchmark, we first consider the impact of PA (insufficiently active [1], regularly active [2]) in the first two lines: note the positive impact (more PA, smaller probability of developing hypertension) in such a group of patients. Next, using 
these two cases as a benchmark, we consider how sex affects the probabilities (the remaining features being the same) in the last column, two first rows; in both cases, women with such a health condition would tend to have a slightly greater probability of developing hypertension, although PA 
would still be beneficial.


Similarly, obesity (next two lines) entails a much higher risk of hypertension for both female and male workers than overweight at all PA levels, although PA still seems beneficial. Of course, when we consider the possibility that the person is overweight or obese (in addition to the other conditions), we obtain intermediate results, as shown in the last two rows.

\noindent
\begin{table}[htb!]
\centering
\footnotesize
\begin{tabular}{c|c|c|c}
BMI                                                         & \begin{tabular}[c]{@{}c@{}}Physical \\ activity\end{tabular} & \begin{tabular}[c]{@{}c@{}}Probability\\ Male \end{tabular} & \begin{tabular}[c]{@{}c@{}}Probability\\ Female \end{tabular} \\ \hline
Overw.                                                 & 1                                                            & 25.26                                                           & 26.34                                                             \\
Overw.                                                 & 2                                                            & 19.79                                                           & 20.70                                                             \\ \hline
Obese                                                       & 1                                                            & 45.54                                                           & 46.95                                                             \\
Obese                                                       & 2                                                            & 34.49                                                           & 35.78                                                             \\ \hline
\begin{tabular}[c]{@{}c@{}}Overw., obese\end{tabular} & 1                                                            & 32.85                                                           & 33.82                                                             \\
\begin{tabular}[c]{@{}c@{}}Overw., obese\end{tabular} & 2                                                            & 22.90                                                           & 23.85                                                             \\ \hline
\end{tabular}
\caption{Probability of developing hypertension given various patient conditions for age greater than 44, poor sleeping level and anxiety.}
\label{tab:Prob_Queries}
\end{table}

 
 \subsection{Health research through hypothesizing evidence}
We can also hypothesize evidence to address various types of health research issues. As an example, let us consider the impact of socioeconomic status on health conditions, as reflected in Table 5. The last column shows the prevalence of hypertension given the three socioeconomic levels: 14.65\% (level 3), 14.90\% (level 2) and 15.48\% (level 1). This suggests a slightly increased prevalence of hypertension as the socioeconomic status decreases.
  \begin{table}[!htb]
    \centering
    \footnotesize
    \begin{tabular}{cccccc}
    Status & Anxiety  & Depression  & Diabetes & Hypercholesterolemia  & Hypertension \\ \hline
   3 &  2.75 & 0.45 & 2.38   &  29.71 & 14.65   \\
    2 &  2.70  & 0.49  & 2.49 & 30.61 & 14.90  \\
    1 & 2.61  & 0.49   &  2.62  & 31.03  & 15.48  \\ \hline
   \end{tabular}
    \caption{Probability of health conditions given socio-economic status.}
    \label{superpedo}
\end{table}
Similarly, we detect a slightly higher prevalence of hypercholesterolemia and diabetes with lower socioeconomic status. By contrast, higher socioeconomic status entails a slightly higher prevalence of anxiety. No effect is noted for depression.

 Table \ref{superpedo2}  describes the BMI population structure depending on the socioeconomic status. For example, for the upper status 3, the structure is 1.18\% for underweight, 42.53\% for normal, 39.63\% for overweight, and 16.66\% for obese.

  \begin{table}[!htb]
    \centering
    \footnotesize
    \begin{tabular}{ccccc}
    Status & Underweight  & Normal & Overweight & Obese \\ \hline
   3 &  1.18 & 42.53   & 39.63   & 16.66    \\
    2 & 1.22  &  41.54  & 40.09  & 17.15    \\
    1 &  1.10 &  38.70  & 41.01  & 19.19    \\ \hline
   \end{tabular}
    \caption{Probability of BMI population structure given socio-economic status.}
    \label{superpedo2}
\end{table}
\noindent  Overall, Table \ref{superpedo2} suggests a slightly worse BMI structure for the lower-income population.
  
\subsection{Therapies through influential findings}

Once we have determined a case of interest (group or individual) and assessed the relevant probability, we may wish to ascertain the influential findings over such statement. For this, we need only to eliminate the conditioning events one-by-one and determine the impact on the target probability, assessing the most influential finding based on the greatest reduction in probability. This is especially relevant in connection to modifiable CVRFs, for which we could explore the best modifications to suggest and communicate interventions to the individual or group by identifying those leading to the most significant reductions.

Consider a base case of a male individual aged 44–54 with  education and socioeconomic status 3, who is obese, inactive, non-smoker, short sleep duration, and has anxiety but no depression. His probability of developing hypertension is 45.63\%. Table 7 shows the probabilities when one of the included modifiable CVRFs is improved to its normal level and the others remain above normal. For example, the first row reflects that the probability of that individual having hypertension if we can improve his BMI level to normal is 11.30\%.

 \begin{table}[!htb]
    \centering
    \footnotesize
    \begin{tabular}{ccc}
    MCVRF & Level & Probability   \\ \hline
   BMI & Normal &  11.30    \\
   Physical activity & Regularly active &  34.57   \\
   Sleep & Normal & 39.69 \\
   Anxiety & No  &  37.02   \\ \hline
   \end{tabular}
    \caption{Probability of hypertension given improved conditions
 for a male aged 44-54, education 
and socioeconomic status 3, obese, low PA, insufficient sleep,
with anxiety but no depression.}
    \label{superkaka}
\end{table}

\noindent Consequently, it seems that the most effective intervention would be to recommend a diet to bring the BMI back to normal. Should we be able to improve the four conditions (i.e., reduce BMI, increase PA and sleep duration, while eliminating anxiety), the probability is reduced to 4.80\%.

Note though, that, in order to elect the best 
recommendations, we would
need to consider the possible impacts (e.g., costs, comfort, quality of life) of the medical conditions and treatments, possibly through a utility function and 
expected utility calculations \citep{French}.


\section{Discussion}\label{sec:discussion}

We have built a discrete BN that relates modifiable and non-modifiable CVRFs and relevant medical conditions and illustrated its use to forecast various conditions, support health research, determine relevant findings and facilitate treatment and policy choices. We have acknowledged uncertainty in the probability tables through posterior distributions.

Once the network is built (i.e., its structure and probability tables), implementing its computations is relatively straightforward using software like GeNIe or R. As an example, the appendix describes one version in GeNIe. Based on this, we could build a decision support system to facilitate clinical and policy decisions.

Two dynamic aspects not reflected in our model are relevant. First, additional data will be acquired annually by the insurance provider and the probability tables will be updated using Bayes’ formula. Second, we have aggregated the data over the years because of homogeneity, as described in Section \ref{sec:datase}; it would be interesting to develop a dynamic BN model to test for health status changes over time.

Finally, for more coherent decision support in choosing the best recommendation, we would need to consider the possible overall impact of the medical conditions and treatments
 using utility functions. In such cases, we would find the portfolio of recommendations with maximum expected utility.

\appendix
\subsection*{Appendix: A GeNIe implementation of the model}
\begin{figure}[H]
\includegraphics[width=14cm]{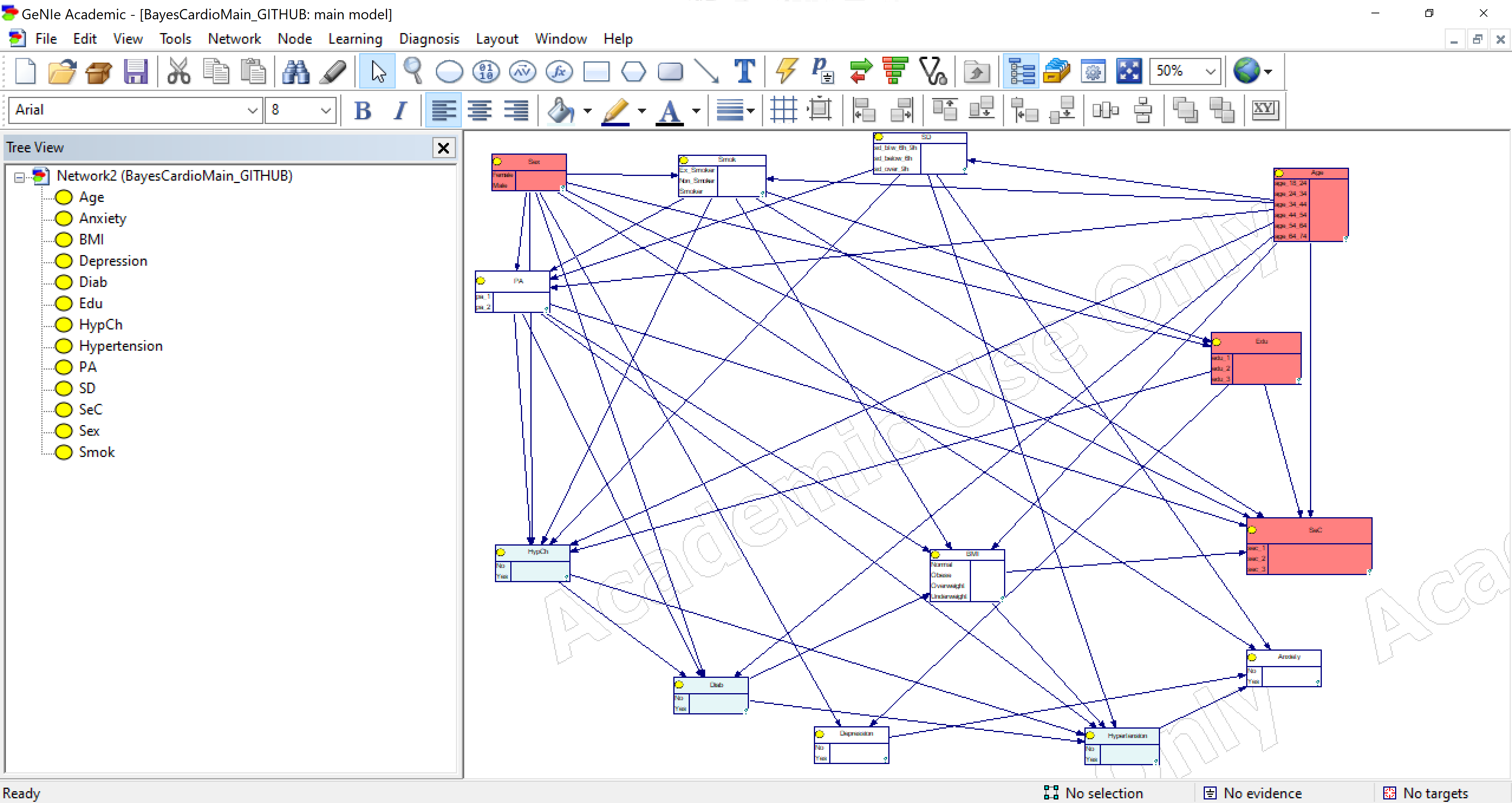}
\centering
\caption{Interface of our model GeNIe implementation.}
\label{fig:interface}
\end{figure}

A GeNIe version of the model (only for academic use) is available from \url{https://github.com/jmcamachor1/CVD_GeNIe_network}.
Figure \ref{fig:interface} displays the interface of the software,
which requires prior downloading of GeNIe for proper 
functioning.

\noindent Based on it, we need only to use the GeNIe features to set evidence at nodes and declare other
nodes as targets to find the required conditional probabilities.  Clearing such evidence and targets reinitiates the original model and habilitates new probabilistic queries to the network. In such a way, we have  robust 
interactive means to undertake research based on the 
proposed model.

\paragraph{\textbf{Declaration of competing interest}}
We confirm that there are not conflict of interest associated with this publication.

\paragraph{\textbf{Acknowledgements}}
Research supported by the AXA-ICMAT Chair in Adversarial Risk Analysis and the Spanish Ministry of Science program MTM2017-86875- C3-1-R. 
We are grateful to Quirónprevención for the provision of data.
Research by A. Lucia is funded by the Spanish Ministry of 
Economy and Competitiveness and FEDER funds from the 
European Union (Grant P/18/00139). Discussions with Victoria 
Ley are gratefully acknowledged.
\bibliographystyle{elsarticle-num} 
\bibliography{cas-refs}

\end{document}